\documentclass[%
 aip,
 amsmath,amssymb,
 reprint,%
]{revtex4-1}

\usepackage{graphicx}
\usepackage{dcolumn}
\usepackage{bm}
\usepackage{color}

\usepackage[utf8]{inputenc}
\usepackage[T1]{fontenc}
\usepackage{mathptmx}
\UseRawInputEncoding

\begin{document}

\preprint{AIP/123-QED}

\title{A simple method to find temporal overlap between THz and X-ray pulses using X-ray-induced carrier dynamics in semiconductors}

\author{Yuya Kubota}
\email{kubota@spring8.or.jp}
\affiliation{RIKEN SPring-8 Center, 1-1-1 Kouto, Sayo, Hyogo 679-5148, Japan}

\author{Takeshi Suzuki}%
\affiliation{Institute for Solid State Physics, The University of Tokyo, Kashiwa, Chiba 277-8581, Japan}

\author{Shigeki Owada}
\affiliation{Japan Synchrotron Radiation Research Institute (JASRI), 1-1-1 Kouto, Sayo, Hyogo 679-5198, Japan}
\affiliation{RIKEN SPring-8 Center, 1-1-1 Kouto, Sayo, Hyogo 679-5148, Japan}

\author{Kenji Tamasaku}
\affiliation{RIKEN SPring-8 Center, 1-1-1 Kouto, Sayo, Hyogo 679-5148, Japan}

\author{Hitoshi Osawa}
\affiliation{Japan Synchrotron Radiation Research Institute (JASRI), 1-1-1 Kouto, Sayo, Hyogo 679-5198, Japan}

\author{Tadashi Togashi}
\affiliation{Japan Synchrotron Radiation Research Institute (JASRI), 1-1-1 Kouto, Sayo, Hyogo 679-5198, Japan}
\affiliation{RIKEN SPring-8 Center, 1-1-1 Kouto, Sayo, Hyogo 679-5148, Japan}

\author{Kozo Okazaki}
\affiliation{Institute for Solid State Physics, The University of Tokyo, Kashiwa, Chiba 277-8581, Japan}

\author{Makina Yabashi}
\affiliation{RIKEN SPring-8 Center, 1-1-1 Kouto, Sayo, Hyogo 679-5148, Japan}
\affiliation{Japan Synchrotron Radiation Research Institute (JASRI), 1-1-1 Kouto, Sayo, Hyogo 679-5198, Japan}

\date{\today}

\begin{abstract}
X-ray-induced carrier dynamics in silicon and gallium arsenide were investigated through intensity variations of transmitted terahertz (THz) pulses in the pico to microsecond time scale with X-ray free-electron laser and synchrotron radiation.
We observed a steep reduction in THz transmission with a picosecond scale due to the X-ray-induced carrier generation, followed by a recovery on a nano to microsecond scale caused by the recombination of carriers.
The rapid response in the former process is applicable to a direct determination of temporal overlap between THz and X-ray pulses for THz pump-X-ray probe experiments with an accuracy of a few picoseconds.
\end{abstract}

\maketitle

%

The development of lasers with ultrashort pulses has revealed transition phenomena to exotic states in condensed matter that cannot be achieved under thermal equilibrium~\cite{Basov2017}.
Recently, the technological advancement in the terahertz (THz) frequency range has opened up new horizons in this field~\cite{Kampfrath2013}.
THz light enables resonant excitation in fundamental phenomena, such as the oscillations of lattices and the precessions of spins.
Furthermore, THz laser pulses can be used as a driving source of intense electric and/or magnetic fields in the temporal form of mono to a few optical cycles.
On the other hand, light with a short wavelength in the X-ray region is very useful as a probe to obtain microscopic properties in electronic and lattice systems.
A combination of these distinctive light sources, a THz pump-X-ray probe method, is a powerful technique, which has been used to investigate lattice and superlattice dynamics after resonant excitations of quasiparticles such as phonon and collective modes of polar vortices~\cite{Kozina2017, Kozina2019, Li2021}.
In addition to THz pulses generated using optical lasers, accelerator-based THz sources have been developed for THz pump-X-ray probe experiments~\cite{Pan2019, Zapolnova2020, Zhang2020, Kang2023}.

Practically, the determination of temporal overlap between THz and X-ray pulses is a technical challenge due to the lack of detectors applicable to both wavelength ranges.
For this purpose, near-infrared (NIR) pulses have been frequently used as an intermediary between THz and X-ray pulses.
For example, the temporal overlap between the X-ray and NIR laser pulses is determined using reference samples, such as yttrium aluminum garnet (YAG) and bismuth~\cite{Sato2019, Kubota2023}, followed by a determination of the temporal overlap between the THz and NIR laser pulses with the electro-optic (EO) sampling method~\cite{Wu1995}.
However, this method is quite complex in terms of optics construction and procedures.
A simple method for determining the temporal overlap with a picosecond accuracy, without using the intermediate NIR pulses, is desirable to efficiently perform the THz pump-X-ray probe experiments.

For this purpose, we investigate the applicability of an X-ray pump-THz probe method.
So far, an X-ray pump-NIR probe method helps determine the temporal overlap between X-ray and NIR pulses using the change in transmission of NIR light in materials, such as YAG and gallium arsenide (GaAs), excited with intense X-rays~\cite{Harmand2013, Sato2015, Chollet2015, Sato2019}.
X-ray absorption excites electrons, which are then relaxed through secondary processes such as Auger decays and electron-electron scattering.
A number of carriers generated in this process changes the sample's refractive index, which is not limited to the NIR region but can be detectable with THz light~\cite{Ulbricht2011, Zapolnova2020}.

In this study, X-ray pump-THz probe experiments were performed at SPring-8 BL19LXU~\cite{Yabashi2001} and SACLA BL3~\cite{Ishikawa2012}.
The X-ray-induced carriers in semiconductors were detected with THz pulses, and the temporal response was investigated in the pico to microsecond time scale.

\begin{figure}
\begin{center}
\includegraphics[width=8cm]{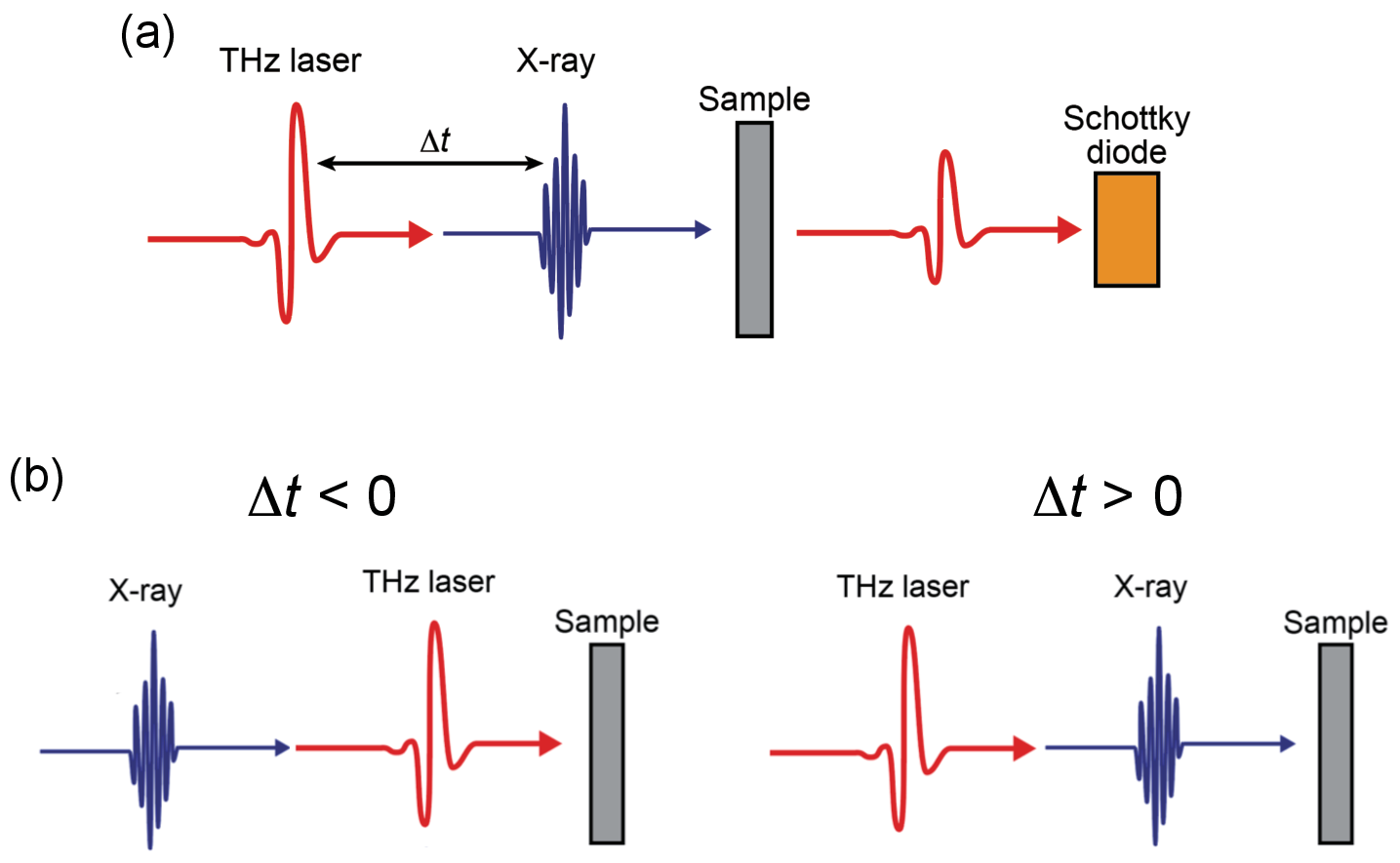}
\caption{
(a) Schematic of the X-ray pump-THz probe experiment for semiconductors.
(b) Schematic of the X-ray and THz pulses reaching the sample. At positive (negative) delay times, the X-ray (THz) pulse reaches the sample first.
}
\label{Fig1}
\end{center}
\end{figure}
Figure~\ref{Fig1}(a) shows a schematic drawing of the X-ray pump-THz probe experiment setup. Silicon (Si) wafers with the thickness of $200$, $500$, and $1000$~$\mu$m, and a GaAs wafer with the thickness of $700$~$\mu$m were used as semiconductor samples, which have been extensively studied using an optical pump-THz probe method~\cite{Suzuki2009, Suzuki2011, Suzuki2012, Terashige2015, Huber2001, Kaindl2003, Huber2005, Kaindl2009, Sekiguchi2017}.
The X-ray-induced carrier dynamics were investigated through intensity variations of the transmitted THz pulses detected with a Schottky diode.
The signal was amplified and gated with a Boxcar integrator.
In this study, a delay time ($\Delta t$) is defined relative to the X-ray (pump) pulse, as shown in Fig.~\ref{Fig1}(b).

\begin{figure*}
\begin{center}
\includegraphics[width=18cm]{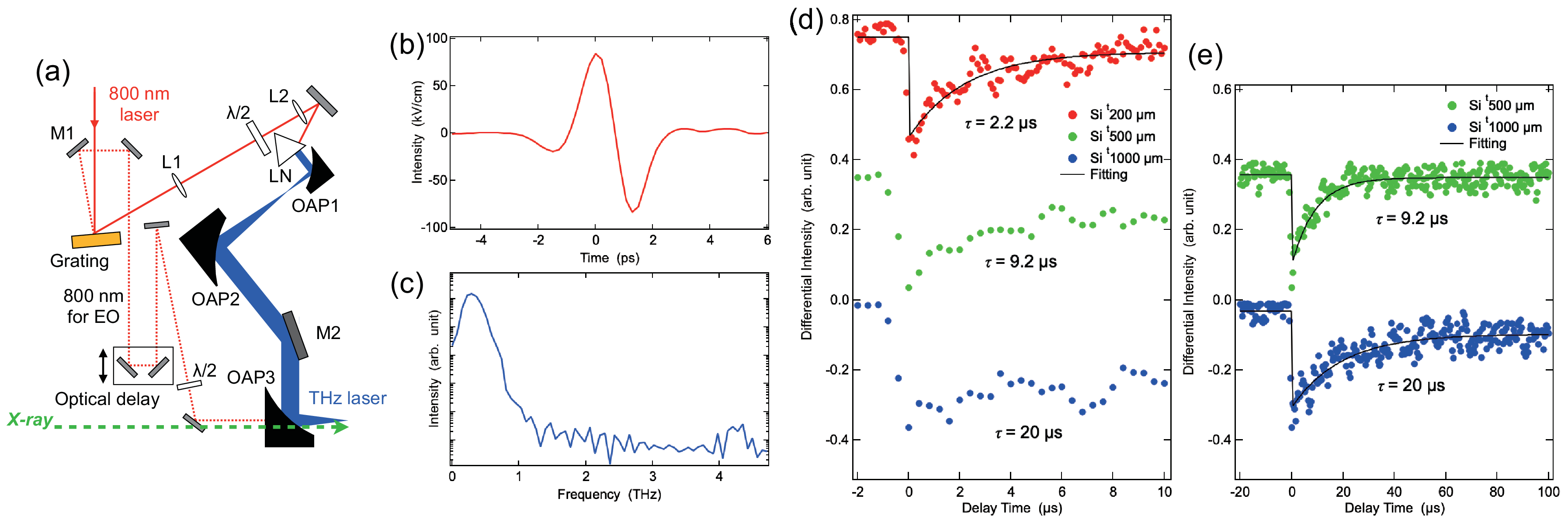}
\caption{
(a) Schematic of the THz pulse generation setup at SPring-8 BL19LXU.
Red solid line: optical path for $800$~nm beam, red dashed line: optical path for THz diagnostic, blue bold line: optical path for THz beam.
Grating: groove density of $1200$~mm$^{-1}$, L1, L2: lenses, LN: LiNbO$_3$ crystal, M1: dielectric coated plane mirror, M2: gold-coated plane mirror,  OAP: Off-axis parabolic mirror.
(b, c) The waveform of the THz pulse evaluated with the EO sampling method (b) and its Fourier spectrum (c).
(d, e) Time evolution of the differential intensity of transmitted THz pulses between the pumped and unpumped conditions for Si with the delay range of (d) [$-2$~$\mu$s, $10$~$\mu$s] and (e) [$-20$~$\mu$s, $100$~$\mu$s].
The red, green, and blue circles represent data for Si with the thickness of $200$, $500$, and $1000$~$\mu$m, respectively.
The black solid lines represent fitting curves with an exponential decay function convoluted with a Gaussian function.
The curves are offset for clarity.
}
\label{Fig2}
\end{center}
\end{figure*}
Figure~\ref{Fig2}(a) shows a schematic drawing of the THz pulse generation setup using the wavefront-tilted scheme~\cite{Hebling2002} at SPring-8 BL19LXU.
We used a Ti:sapphire-based regenerative amplifier system with a pulse energy of $\sim0.5$~mJ, a pulse duration of $120$~fs, a repetition rate of $9$~kHz, and a central wavelength of $800$~nm.
A grating with a groove density of $1200$~cm$^{-1}$ was used to tilt the wavefront.
The incident and diffracted angle of the grating were $4.9$~degrees and $60.9$~degrees, respectively.
The reflected beam was used to detect the THz waveform using the EO sampling method.
The image of the grating was transferred to the LiNbO$_3$ (LN) crystal using L1 and L2 lenses located in the $4f$ configuration.
The generated THz pulses were expanded, collimated, and focused using off-axis parabolic (OAP) mirrors.
Figures~\ref{Fig2}(b) and \ref{Fig2}(c) show the waveform of the THz pulse evaluated by the EO sampling method with a 300-$\mu$m-thick GaP crystal and its Fourier spectrum, respectively.

Figures~\ref{Fig2}(d) and \ref{Fig2}(e) show the time evolution of the differential intensity of transmitted THz pulses between the pumped and unpumped conditions for Si.
The samples were excited by X-ray pulses with a repetition rate of $4.5$~kHz divided by a chopper.
The photon energy $h\rm \nu$, the photon flux, and the size of the X-ray were $10$~keV, $5.5 \times 10^6$~photons/pulse, and $500$~$\mu$m $\times$ $500$~$\mu$m, respectively.
The temporal duration of the X-ray pulse was $\sim 40$~ps, which dominantly determined the total time resolution.
Note that we confirmed that the Schottky diode did not react to X-ray pulses transmitted through the sample.
As seen in Figs.~\ref{Fig2}(d) and \ref{Fig2}(e), we observed a rapid decrease in the THz transmission at $\Delta t = 0$, followed by a slow recovery.
This behavior suggests the fast generation of electron-hole pairs by the X-ray pulse and the slow recombination of the pairs.
These results indicate that the X-ray-induced carriers are detected with THz pulses.
The relaxation time of the transmission ($\tau$) was evaluated to be $2.2$, $9.2$, and $20$~$\mu$s for Si with the thickness of $200$, $500$, and $1000$~$\mu$m, respectively, by fitting using an exponential decay function convoluted with a Gaussian function.
These values are comparable to the lifetime of free carriers in Si~\cite{Schroder1997}.

\begin{figure}
\begin{center}
\includegraphics[width=7cm]{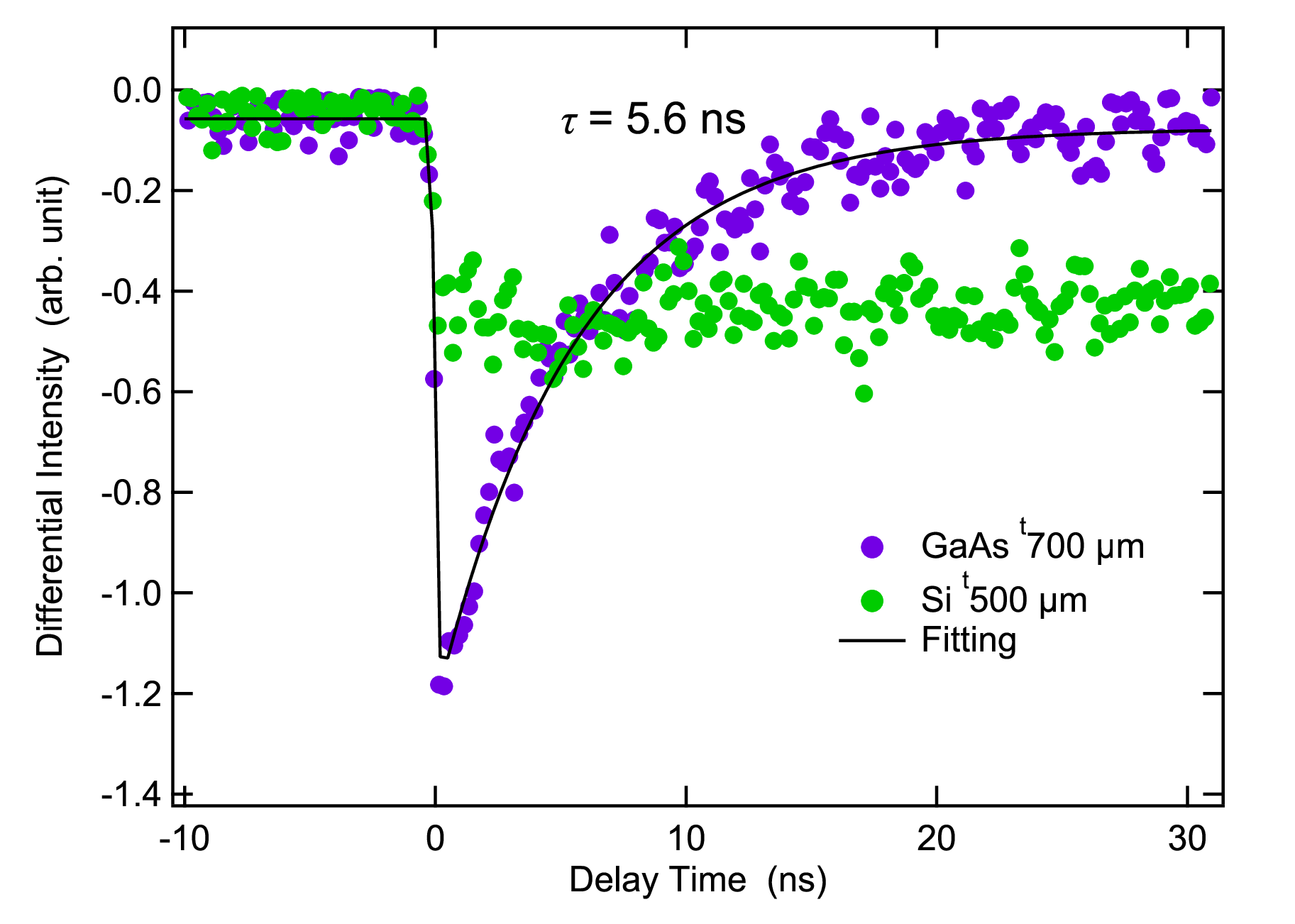}
\caption{
Time evolution of the differential intensity of transmitted THz pulses between the pumped and unpumped conditions for GaAs with the thickness of $700$~$\mu$m (purple circles).
For comparison, the data for Si with the thickness of $500$~$\mu$m are also plotted (green circles).
The black solid line represents a fitting curve with an exponential decay function convoluted with a Gaussian function.
}
\label{Fig3}
\end{center}
\end{figure}
In contrast to the results on Si, the response of GaAs has distinctive characteristics due to its unique electronic band structure.
Figure~\ref{Fig3} shows the time evolution of the transmission of THz pulses for GaAs.
The relaxation time constant for GaAs is estimated to be 5.6 ns, determined using the same fitting procedure as for Si.
This value is significantly shorter than those of Si and agrees with previous reports~\cite{Hooft1987, Lai2006}.
This difference is attributed to GaAs having a direct band gap structure, where electrons and holes can recombine efficiently by emitting light.
In contrast, Si has an indirect band gap structure, which makes recombination less efficient and results in a longer relaxation time.
We will discuss the carrier dynamics in detail later.

\begin{figure*}
\begin{center}
\includegraphics[width=18cm]{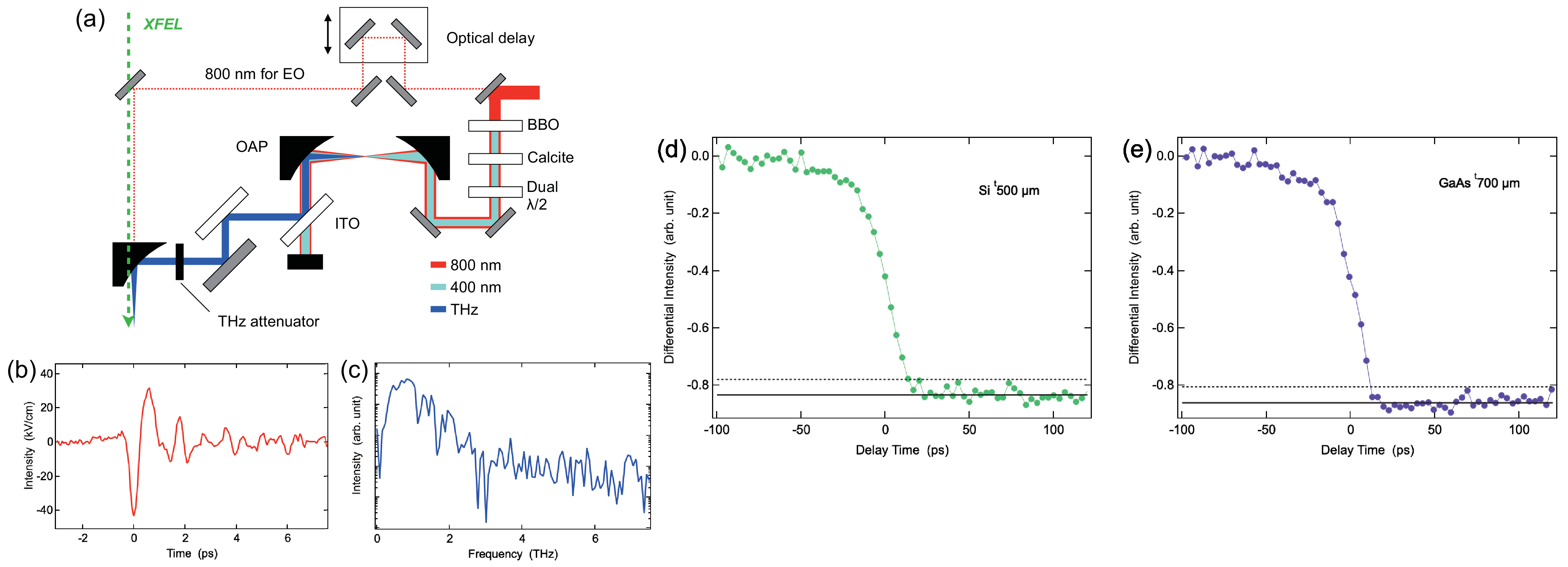}
\caption{
(a) Schematic of the THz pulse generation setup at SACLA BL3.
Red bold line: optical path for $800$~nm beam, red dashed line: optical path for THz diagnostic, cyan bold line: optical path for $400$~nm beam, blue bold line: optical path for THz beam.
BBO: Beta Barium Borate crystal, ITO: indium tin oxide coated mirror, OAP: Off-axis parabolic mirror.
(b, c) The waveform of the THz pulse evaluated with the EO sampling method (b) and its Fourier spectrum (c).
(d, e) Time evolution of the differential intensity of transmitted THz pulses between the pumped and unpumped conditions for Si with the thickness of $500$~$\mu$m (d) and GaAs with the thickness of $700$~$\mu$m (e).
Delay time zero is defined as the center of the falling.
Black solid and dashed lines represent the average and $3 \sigma$ values in the $20$ to $170$~ps range, respectively.
}
\label{Fig4}
\end{center}
\end{figure*}
To observe the carrier dynamics on the picosecond timescale, we performed an experiment at SACLA BL3 using X-ray free-electron laser (XFEL) pulses with a temporal duration of less than $10$~fs~\cite{Inubushi2012, Inoue2019, Osaka2022}.
For SACLA BL3, we adopted a method with two-color laser-induced gas plasma~\cite{Cook2000} to simply generate intense THz pulses, as shown in Fig.~\ref{Fig4}(a).
This is because a shorter pulse duration of $40$~fs and higher pulse energy of $\sim 12$~mJ for the NIR laser with a Ti:sapphire-based regenerative amplifier system is available~\cite{Togashi2020}.
The repetition rate and the central wavelength of the NIR laser were $30$~Hz and $800$~nm, respectively.
A second harmonic pulse at $400$~nm was generated with a Beta Barium Borate (BBO) crystal.
The fundamental and second harmonic pulses interacted at a focal point of an OAP mirror and generated a THz pulse.
The THz pulses were collimated and focused using other OAP mirrors.
Figures~\ref{Fig4}(b) and \ref{Fig4}(c) show the waveform of the THz pulse evaluated with the same method as SPring-8 and its Fourier spectrum, respectively.

Figure~\ref{Fig4}(d) shows the time evolution of the differential intensity of transmitted THz pulses between the pumped and unpumped conditions for Si with the thickness of $500$~$\mu$m.
The photon energy $h\rm \nu$, the pulse energy, the size, and the repetition rate of XFEL were $10$~keV, $1.2 \times 10^9$~photons/pulse, $344$~$\mu$m $\times$ $368$~$\mu$m, and $15$~Hz, respectively.
To prevent the XFEL pulses transmitted from the sample from affecting the Schottky diode, a gold mirror was placed downstream of the sample to reflect and detect only the THz pulses.
We observed a decrease in the THz transmission due to X-ray-induced carriers, which is consistent with the results obtained at SPring-8.
The fall time of the transmission is $\sim 20$~ps, corresponding to the temporal duration of the THz pulse.
Note that the chirped pulse induced by the carriers in the sample may also contribute to the fall time.
Furthermore, we observed an asymmetry shape in the fall of the transmission, where it shows a gradual change in the negative delay range ($\Delta t < 0$) while a steep reduction in the positive delay range ($\Delta t > 0$).
This response is explained by the temporal profile of the THz pulse.
The THz pulse generated by the gas-plasma method has a main peak at the beginning of the pulse and a long tail, as depicted in Fig.~\ref{Fig4}(b).
In $\Delta t < 0$, the X-ray pulse overlaps with the tail end of the THz pulse, resulting in a gradual change in the THz transmission.
In contrast, in $\Delta t > 0$, the X-ray pulse overlaps with the main peak of the THz pulse, and the transmission exhibits a steep change.
If we define the beginning of the THz pulse as the time when the intensity changes by more than $3 \sigma$ from the averaged value in the positive delay range above $20$~ps, we were able to determine the temporal overlap with an accuracy of less than $2$~ps, as shown in Figs.~\ref{Fig4}(d).
This value is a sufficient accuracy to find the overlap from a wide temporal range in the THz pump-X-ray probe experiments.
The tail of the THz pulse is generated by vibrations of water in the air. The temporal resolution of this method may be improved by removing the water on the THz path and shortening the pulse duration.
We note that Fig.~\ref{Fig4}(e) shows the time evolution of the transmission of THz pulses for GaAs.
The response of THz transmission in GaAs shows the same behavior as in Si.
This is because the intensity of transmitted THz pulses was reduced to near the detection limit due to the high XFEL pulse energy, and there is no significant difference in the carrier dynamics between the samples in this time scale.
Note that we also confirmed a decrease in the THz transmission using a smaller XFEL beam size of $5.0$~$\mu$m $\times 7.6$~$\mu$m focused with compound refractive lenses~\cite{Katayama2019}, which is often used for optical pump-XFEL probe experiments at SACLA, for Si with the thickness of 1000~$\mu$m (not shown).

\begin{table}
\caption{Plasma frequencies of Si and GaAs, and parameters to evaluate them.}
\begin{ruledtabular}
\begin{tabular}{lll}
 & Si & GaAs\\
\hline
Quantum efficiency $Z_0$ & $2.7 \times 10^3$ & $2.4 \times 10^3$\\
Penetration depth $d$ & $133.71$~~$\mu$m & $51.78$~~$\mu$m\\
Reduced mass $m^{\ast}$ & $0.123 m_0$ & $0.044 m_0$\\
Plasma frequency $f_{\mathrm{pl}}$ & $0.73$~THz & $1.8$~THz\\
\end{tabular}
\end{ruledtabular}
\label{table1}
\end{table}
We briefly discuss the difference in the THz transmission between Si and GaAs obtained at SPring-8 BL19LXU.
The decrease in the transmission through GaAs is more pronounced than that through Si as shown in Fig.~\ref{Fig3}.
To investigate this observation, we performed calculations to determine the plasma frequencies induced by X-ray pulses in both Si and GaAs.
Table~\ref{table1} summarizes the calculations.
Carrier density, $N$, excited by an X-ray pulse is calculated by
\begin{equation}
N = Z_0 J/x y d,
\label{carrier}
\end{equation}
where $Z_0$ and $J$ are quantum efficiency and photon flux, respectively.
We approximate that all X-ray pulse with the size of $x \times y$ is uniformly absorbed within the penetration depth, $d$.
For the X-ray pulse with $10$~keV used in this study, the value of $Z_0$ is $2.7 \times 10^3$ $\left(2.4 \times 10^3 \right)$~\cite{Bertuccio2002}, and $d$ is $133.71$ $\left(51.78 \right)$~$\mu$m~\cite{CXRO} for Si (GaAs).
Using the values of $N$, the plasma frequency $\omega_{\mathrm{pl}}$ is calculated by
\begin{equation}
\omega_{\mathrm{pl}} = 2 \pi f_{\mathrm{pl}} = \sqrt{\frac{N e^2}{\varepsilon_0 m^{\ast}}},
\label{plasma}
\end{equation}
where $e$ and $\varepsilon_0$ are electron charge magnitude and permittivity of free space, respectively.
The values of the reduced mass of electron-hole ($e$-$h$) pairs, $m^{\ast} = \left( 1 / m_e + 1 / m_h \right)^{-1}$, are $0.123 m_0$ for Si~\cite{Lipari1971} and $0.044 m_0$ for GaAs~\cite{Sekiguchi2017}, where $m_0$ is the bare electron mass.
The values of $f_{\mathrm{pl}}$ are evaluated to be $0.73$~THz for Si and $1.8$~THz for GaAs in the situation of the experiment at SPring-8 BL19LXU.
Because $f_{\mathrm{pl}}$ of GaAs is higher than that of Si, GaAs absorbs and reflects the higher frequency components of THz pulses more effectively than Si.
This difference leads to a greater reduction in transmission through GaAs than Si.
Note that the value of $f_{\mathrm{pl}}$ obtained at SACLA BL3 is evaluated to be more than $10$~THz for both Si and GaAs.
This value indicates that the samples absorb and reflect most of the frequency components of the THz pulses resulting in a large reduction of the transmission.

Figures~\ref{Fig2}(d) and \ref{Fig2}(e) show a sample thickness dependence of the relaxation time in Si.
We have also discussed this phenomenon using a simulation with a diffusion equation as shown in the supplementary material.

In summary, the X-ray pump-THz probe experiments for Si and GaAs were performed to establish the simple method to determine the timing overlap between THz and X-ray pulses.
The X-ray-induced carrier dynamics were investigated in a wide temporal range from pico to microsecond with XFEL and synchrotron radiation X-ray pulses.
The difference in carrier dynamics between Si and GaAs was discussed with their band structures and plasma frequencies.
From the presented results, we have confirmed that the method is applicable to the finding of the temporal overlap with an accuracy of a few picoseconds.
Due to its simplicity, this method can be adapted to various THz pump-X-ray probe experiments without any changes in the setup.

See the supplementary material for the simulation with a diffusion equation.

This work was supported by Grants-in-Aid for Scientific Research (KAKENHI) (Grants No. JP24K21043, No. JP24K01375, No. JP24K00565, and No. JP24KF0021) from the Japan Society for the Promotion of Science (JSPS).
T. S. acknowledges research grants from Izumi Science and Technology Foundation. 
This experiment was performed at SPring-8 BL19LXU with the approval of RIKEN (Proposal No. 20210050) and SACLA BL3 with the approval of the Japan Synchrotron Radiation Research Institute (JASRI) (Proposal Nos. 2019B8045 and 2022B8073).

The data that support the findings of this study are available from the corresponding author upon reasonable request.


\end{document}